\documentclass{nature}
\usepackage{hyperref}

\usepackage[utf8]{inputenc}

\usepackage{graphicx}
\makeatletter
\let\saved@includegraphics\includegraphics
\AtBeginDocument{\let\includegraphics\saved@includegraphics}
\renewenvironment*{figure}{\@float{figure}}{\end@float}
\makeatother
\usepackage{float}

\usepackage{caption}
\usepackage{amsmath}
\usepackage{textcomp}
\usepackage{gensymb}
\usepackage{bm}
\usepackage[T1]{fontenc}
\usepackage{ae,aecompl}
\usepackage{amssymb}

\usepackage[super]{nth}

\usepackage{journals}

\usepackage{xspace}
\def\SL{\texttt{Starlight}}
\def\ppxf{\texttt{pPXF}}
\def\Ha{$\text{H}\alpha$}
\def\SLs{\SL\ }
\def\ppxfs{\ppxf\ }
\def\Has{\Ha\ }
\def\Om{$\Omega(r)$}
\def\Omp{$\Omega_{\text{P}}(r)$}
\def\ccang{$\delta\theta(r)$}
\def\Oms{\Om\ }
\def\Omps{\Omp\ }
\def\ccangs{\ccang\ }

\title{A direct test of density wave theory in a grand design spiral galaxy}
\author{Thomas G. Peterken$^1$, Michael R. Merrifield$^1$, Alfonso Arag{\'o}n-Salamanca$^1$, Niv Drory$^2$, Coleman M. Krawczyk$^3$, Karen L. Masters$^{4,3}$, Anne-Marie Weijmans$^5$, and Kyle B. Westfall$^6$}
\begin{document}
\maketitle
\label{firstpage}

\begin{affiliations}
 \item School of Physics and Astronomy, The University of Nottingham, University Park, Nottingham NG7 2RD, UK
 \item McDonald Observatory, The University of Texas at Austin, 1 University Station, Austin, TX 78712, USA
 \item Institute of Cosmology \& Gravitation, University of Portsmouth, Dennis Sciama Building, Portsmouth, PO1 3FX, UK
 \item Haverford College, Department of Physics and Astronomy, 370 Lancaster Avenue, Haverford, Pennsylvania 19041, USA
 \item School of Physics and Astronomy, University of St Andrews, North Haugh, St Andrews, KY16 9SS, UK
 \item University of California Observatories, University of California, Santa Cruz, Santa Cruz, CA 95064, USA
\end{affiliations}

\begin{abstract} \label{summary}

The exact nature of the arms of spiral galaxies is still an open question\cite{DobbsBaba14}. It has been widely assumed that spiral arms in galaxies with two distinct symmetrical arms are the products of density waves that propagate around the disk, with the spiral arms being visibly enhanced by the star formation that is triggered as the passing wave compresses gas in the galaxy disk\cite{LinShu64,Bertin14,DobbsBaba14}. Such a persistent wave would propagate with an approximately constant angular speed, its pattern speed $\Omega_{\text{P}}$. The quasi-stationary density wave theory can be tested by measuring this quantity and showing that it does not vary with radius in the galaxy. Unfortunately, this measurement is difficult because $\Omega_{\text{P}}$ is only indirectly connected to observables such as the stellar rotation speed\cite{TremaineWeinberg84,FontBeckman11,Beckman+18}. Here, we use the detailed information on stellar populations of the grand-design spiral galaxy UGC 3825, extracted from spectral mapping, to measure the offset between young stars of a known age and the spiral arm in which they formed, allowing the first direct measure of $\Omega_{\text{P}}$ at a range of radii. The offset in this galaxy is found to be as expected for a pattern speed that varies little with radius, indicating consistency with a quasi-stationary density wave, and lending credence to this new method.
\end{abstract}

The hypothesized existence of such constant pattern speeds emerged from work in the 1960s on self-exciting density waves in stellar and gaseous disks\cite{LinShu64,LinShu66,Kalnajs65}, coupled with the realisation that the prevalence of long-lived large-scale spiral arms in galaxies means that a single global spiral mode dominates the process\cite{Toomre69,Bertin+89}. primarily driven by the gas dynamics of the disk\cite{BertinLin96}.  As such a global density wave propagates coherently around the galaxy triggering star formation where it compresses gas, it should create a quasi-stationary spiral pattern across the face of disk galaxies\cite{Bertin14,Zhang16}.  As time progresses, newly-formed stars will move away from the peak of the density wave at a rate that depends on the relative speeds of stars and wave.  Thus, the offset between such stars and the current location of the density wave provides a direct measure of the speed at which this wave is propagating.

Using such offsets to determine pattern speeds is not new: it has been employed to good effect using the observed offset between the dense molecular gas that is currently being compressed by the spiral wave and young hot stars that formed previously in the spiral arm and have now moved from the peak of the wave\cite{Egusa+04,Tamburro+08,Egusa+09}. The resulting colour gradient across an arm has also been used to the same effect\cite{GonzalezGraham96,PuerariDottori97,MartinezGarcia+09a,MartinezGarcia+13}.  However, in these previous analyses the offset in time between the two phases -- essentially the timescale for star formation -- was also unknown, so had to be solved for simultaneously.  The price paid for deriving this extra parameter was that $\Omega_{\text{P}}$ had to be assumed to be constant with radius, making it less of a test of the quasi-stationary nature of the conventional density wave picture.  However, we have now reached a point where the quality of optical spectroscopy and the associated modelling techniques allow one to extract a stellar population of a specified age from spectral data, so that $\Omega_{\text{P}}$ can be measured as a function of radius to see how constant it really is.  

As a test case, we have selected the galaxy UGC~3825. This isolated system\cite{Verley+07} has a symmetric grand-design structure, which makes it a prime candidate for being the product of a global density wave of internal (rather than tidal) origin\cite{Bertin14,Pettitt+17}.  According to the Galaxy Zoo citizen science project\cite{Hart+16}, it does not contain a bar, which might complicate the interpretation of its spiral structure, and it is at an ideal intermediate angle to the line of sight, allowing us both to identify its spiral structure and to measure the rotational motion of material via the Doppler shift.  It is also one of the targets of the SDSS-IV MaNGA (Mapping Nearby Galaxies at APO) integral-field spectroscopic survey\cite{Bundy+15}, which means that for every point across the face of the galaxy a high-quality spectrum has been obtained.

\begin{figure}
\centering
\includegraphics[width=\columnwidth]{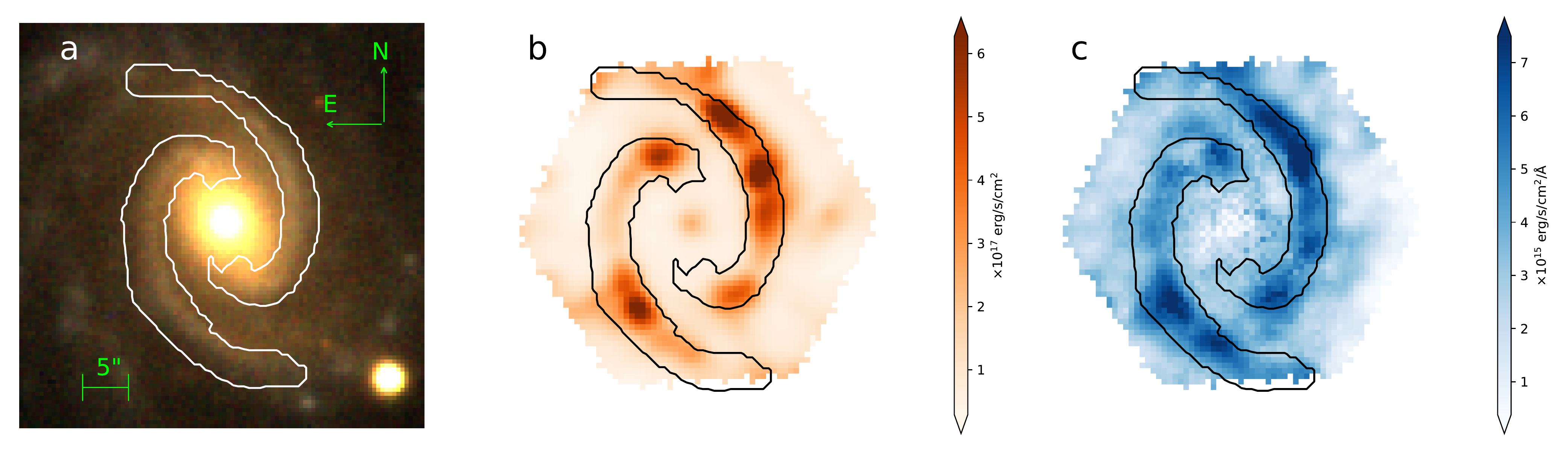}
\caption{\textbf{UGC~3825 and its star formation tracers used here.} (\textit{a}) SDSS imagery\cite{Fukugita+96}, showing 5 arcseconds for scale. (\textit{b}) Ongoing star-formation, traced by the H$\alpha$ emission line. (\textit{c}) 4020 \AA\ flux of young stars. In each panel, the outline denotes the region within which at least 25\% of GZ:3D users agreed on the presence of a spiral arm.}
\label{fig:Populations}
\end{figure}

At each location across the face of the galaxy, we decompose the MaNGA spectrum into the contributions from stars of differing ages\cite{Conroy13} and that from current star formation (see Methods), allowing us to map the distribution of all these various components.  Figure~\ref{fig:Populations} shows the resulting distribution of H$\alpha$ emission and young stars.  The H$\alpha$ emission is a frequently-used indicator for ongoing star formation\cite{Kennicutt98} while the young stars are defined as those with ages up to 60 million years after formation. Since the youngest stars dominate the light, the map picks out the location of the stars at a time $\delta \tau \sim 2 \times 10^{7}$ years after they formed (see Methods), providing us with maps of different timescales since the onset of star formation.  As a fiducial, the figure also shows the location of the spiral arm regions determined as part of the ongoing Galaxy Zoo:3D (GZ:3D) citizen science project (see Methods), although this information on the locations of the arms is not required or used in the analysis here.  Even in these raw maps, it is discernable that over most of the galaxy the young stellar population is found on the leading edge of the spiral arm.  This is what one would expect from the spiral density wave picture, as the material in the inner parts of a galaxy is predicted to circulate at a higher angular velocity than the spiral pattern, so gas clouds overtake the arms and collapse to form stars in the density wave. These young stars continue to overtake the spiral arm to emerge out of the leading edge after a time interval determined by the difference between the pattern and material speeds.

We can render this description more quantitative by measuring the small angular offset in azimuth between these two spiral arm tracers as a function of radius, $\delta\theta(r)$, by cross correlating data from the maps of current star formation and the young stellar population.  We can also measure the circular angular speed of the galaxy as a function of radius, $\Omega(r)$, using the Doppler shift in the emission lines in the spectra (appropriately corrected for the galaxy's inclination; see Methods for details).  By considering the rate at which material travelling around the galaxy at this angular speed will overtake the spiral pattern, it is straightforward to show\cite{Egusa+09} that the pattern speed is given by the formula 
\begin{equation}
\Omega_{\text{P}}(r) = \Omega(r) - \frac{\delta\theta}{\delta \tau}(r).
\label{eq:patternspeed}
\end{equation}

\begin{figure}
\centering
\includegraphics[width=\columnwidth]{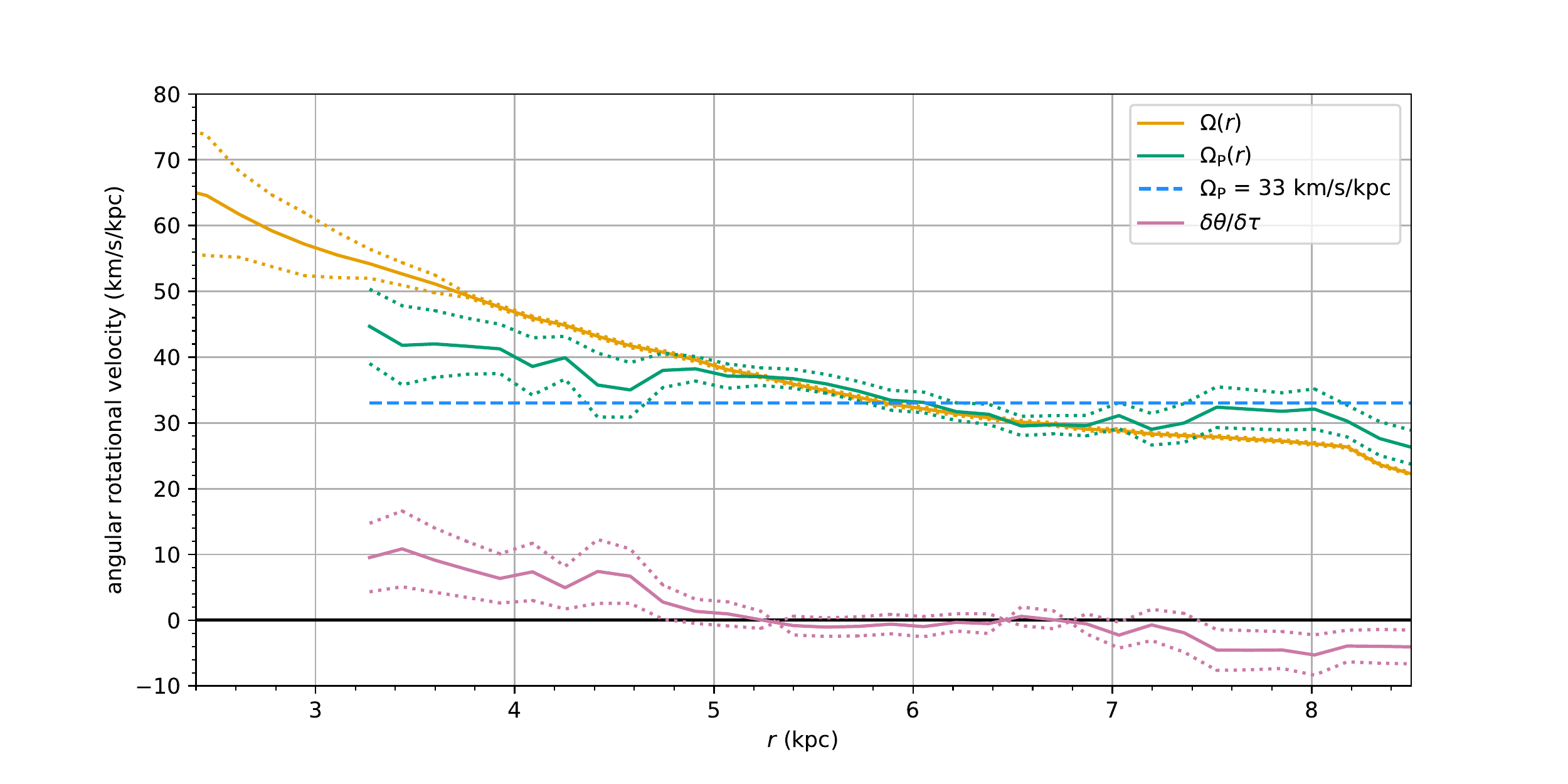}
\caption{\textbf{The derived pattern speed \Omps for UGC~3825 using this method is shown in green.} This is the value of $\frac{\delta\theta}{\delta\tau}(r)$ (plum solid) subtracted from the angular velocity of material \Oms (gold solid), as defined in Equation \ref{eq:patternspeed}. The estimated 1 $\sigma$ uncertainties in these quantities are denoted by dotted lines (see Methods for details). The light blue dashed line shows a best-fit flat line to the pattern speed. The pattern-speed calculation is limited to the region where at least 25\% of the GZ:3D users agree on the location of the spiral arms.}
\label{fig:Pattern_speed}
\end{figure}

The results of this analysis for UGC~3825, presented in Figure~\ref{fig:Pattern_speed}, are entirely consistent with the predictions of the density wave theory.  At small radii, as expected from the qualitative analysis of offsets, matter is rotating faster than the derived pattern speed, but eventually the measured angular speed of material drops to where it is rotating at the same speed as the spiral pattern, a point known as the corotation resonance.  The location of this resonance, at a radius of $\sim 6\,{\rm kpc}$ ($0.6\,R_{\textrm{E}}$), is consistent with estimates for other galaxies using less direct techniques\cite{Tamburro+08} and with the predictions for a modal picture of quasi-stationary density waves\cite{BertinLin96,Bertin14}.  The derived form for $\Omega_{\text{P}}(r)$ is consistent with a constant value of $33\,{\rm km}\,{\rm s}^{-1}{\rm kpc}^{-1}$. Such constancy was in no way imposed by the analysis, but rather again confirms the existence of a quasi-stationary density wave for this galaxy.


Thus, at least for the case of UGC~3825, a coherent story emerges in which the observed spiral structure is consistent with a quasi-stationary internal-origin density wave. However, it has been suggested (and evidence is making it increasingly clear) that such a model can only explain the spiral structure found in a fraction of all galaxies\cite{Meidt+09,Hart+17}.  Since different mechanisms for producing spiral arms should result in significantly different radial profiles in pattern speed\cite{Bertin14,Baba+15}, we can distinguish between such physical processes using this new technique; large spectroscopic surveys of galaxies like MaNGA will ultimately allow us to fully determine the circumstances under which galaxy spiral arms are produced by long-lived density waves.

\appendix

\section{Acknowledgements}


This work makes extensive use of the \SLs and \ppxfs spectral fitting tools, both of which are freely available. The \SL project is available at \url{http://www.starlight.ufsc.br/} and is supported by the Brazilian agencies CNPq, CAPES and FAPESP and by the France-Brazil CAPES/Cofecub program. \ppxf was created and is maintained by Michele Cappellari and is available at \url{http://www-astro.physics.ox.ac.uk/~mxc/software/}. The table-matching tool \texttt{TOPCAT} (available at \url{http://www.star.bris.ac.uk/\%7Embt/topcat/}) was also used in this work.

Several Python tools were also essential for this research. \texttt{Astropy} is a community-developed core Python package for Astronomy available at \url{http://www.astropy.org/}. \texttt{Scipy} is an open-source scientific computing package available at \url{http://www.scipy.org/}. The figures in this Letter were generated using \texttt{matplotlib}, available at \url{https://matplotlib.org/}.


This publication uses data generated via the Zooniverse.org platform, development of which is funded by generous support, including a Global Impact Award from Google, and by a grant from the Alfred P. Sloan Foundation. This publication has been made possible by the participation of almost 6000 volunteers in the Galaxy Zoo:3D project on Zooniverse.org.

Funding for the Sloan Digital Sky Survey IV has been provided by the Alfred P. Sloan Foundation, the U.S. Department of Energy Office of Science, and the Participating Institutions.  SDSS-IV acknowledges support and resources from the Center for High-Performance Computing at the University of Utah. The SDSS web site is \url{www.sdss.org}.

SDSS-IV is managed by the Astrophysical Research Consortium for the Participating Institutions  of  the  SDSS  Collaboration including the Brazilian Participation Group, the Carnegie Institution for Science, Carnegie Mellon University, the Chilean Participation Group, the French Participation Group, Harvard-Smithsonian Center for Astrophysics, Instituto de Astrof{\'i}sica de Canarias, The Johns Hopkins University, Kavli Institute for the Physics and Mathematics of the Universe (IPMU) / University of Tokyo, Lawrence Berkeley National Laboratory, Leibniz Institut f{\"u}r Astrophysik Potsdam (AIP), Max-Planck-Institut f{\"u}r Astronomie (MPIA Heidelberg), Max-Planck-Institut f{\"u}r Astrophysik (MPA Garching), Max-Planck-Institut f{\"u}r Extraterrestrische Physik (MPE), National Astronomical Observatories of China, New Mexico State University, New York University, University of Notre Dame, Observat{\'o}rio Nacional / MCTI, The Ohio State University,  Pennsylvania State University, Shanghai Astronomical Observatory, United Kingdom Participation Group,  Universidad Nacional Aut{\'o}noma de M{\'e}xico, University of Arizona, University of Colorado Boulder, University of Oxford, University of Portsmouth, University of Utah, University of Virginia, University of Washington, University of Wisconsin, Vanderbilt University, and Yale University.

We are grateful for access to the University of Nottingham High Performance Computing Facility, without which the spectral fitting work done here would not have been possible in any reasonable timeframe.

\section{Author contributions}

Correspondence should be directed to T.P.. M.M. conceived of the presented idea. T.P. developed the method and obtained the results. A.A.S. and M.M. supervised the work and led the main analysis and interpretation with T.P.. K.W. calculated the kinematic parameters of UGC 3825 and discussed the implications of the results presented here. K.M. and C.K. devised, implemented, and provided output from the Galaxy Zoo:3D spiral arm mask project. N.D., K.M., A.W., K.W., and many others in the SDSS community obtained the MaNGA IFU data, developed reduction and analysis code, continue to maintain software and hardware, and performed many other tasks necessary for the running of a large collaboration. All authors discussed the results and contributed to the final manuscript.

\section{Competing Interests}
The authors declare that they have no competing financial interests.

\section{Author Information}
Correspondence should be directed to T.P. (Thomas.Peterken@nottingham.ac.uk).

\section{Methods}

\subsection{MaNGA}
The SDSS-IV MaNGA survey\cite{Blanton+17,Yan+16-design} provides integral-field spectroscopy\cite{Drory+15,Law+15,Law+16} for a large sample of low-redshift galaxies\cite{Wake+17} using the BOSS spectrograph\cite{Smee+13,Yan+16-cal} on the SDSS telescope\cite{Gunn+06} located at Apache Point, New Mexico. Please refer to referenced publications on the technical specifications of the survey\cite{Blanton+17,Yan+16-design,Law+16,Law+15,Wake+17,Smee+13,Yan+16-cal,Gunn+06,Drory+15}.

This work makes use of the IFS mapping of UGC~3825 (MaNGA plate-IFU 8132-12702), which is publicly available as part of the fourteenth SDSS data release (DR14)\cite{SDSSDR14}. As part of this work, we make use of the MaNGA data analysis pipeline (DAP) outputs as part of the internal data release MPL-6. These DAP outputs will be publicly available as part of SDSS DR15, planned for December 2018.

\subsection{Galaxy Zoo:3D}
Figure~1 makes use of spiral arm masks generated by Galaxy~Zoo:3D (GZ:3D) as a fiducial. This is an ongoing citizen science project at \url{https://www.zooniverse.org/projects/klmasters/galaxy-zoo-3d} inspired by the Galaxy~Zoo\cite{Lintott+08,Lintott+11} galaxy classification project. In GZ:3D, volunteer users are shown images of galaxies in the MaNGA target catalogue and are asked to draw boundaries marking the edges of spiral arms and bars, as well as identifying the position of the galactic centre. The end result is an image with defined spiral and bar weights at each position, determined by the number of users who defined that position as part of a spiral arm or a bar. Further details will be published in a separate paper by Masters et al. (in preparation). For UGC~3825, 8 users drew spiral arms, so the fiducial in Figure~1 is a contour denoting the contiguous region where at least 2 users defined as being part of the spiral arms.

As stated in the main text, the GZ:3D spiral arm masks are only used for display purposes in Figure~1, and are not used in the analysis.

\subsection{Spectral fitting} MaNGA data provide a cube of information comprising a spectrum from 3600~\AA\ to 10000~\AA\ at each spatial location on the sky.  Each such spectrum of UGC~3825 (MaNGA plate-IFU 8132-12702) was fitted using a set of 270 single stellar population (SSP) template spectra from the MILES project\cite{Vazdekis+15}. The MILES templates have a wavelength range similar to MaNGA (3540~\AA\ to 7410~\AA), with SSPs available for a large number of different ages and metallicities. We use templates covering a wide range of ages (27 values between  $3\times10^{7}$~($\approx10^{7.5}$)~years and $13\times10^9$~($\approx10^{10.1}$)~years) and metallicities (10 values of [M/H] between -1.79 and +0.40). The finer sampling of the age parameter space was required to achieve the temporal resolution needed to separate out the young stellar components sought in this analysis; the coarser sampling in metallicity is entirely adequate for this work while keeping the total number of templates within the maximum that the software can process. We assume a Kroupa revised stellar initial mass function (IMF)\cite{Kroupa01} with Padova isochrones\cite{Girardi+00}. We fit each spaxel's spectrum individually to ensure that we retain all of the spatial information possible. This will result in decreased signal-to-noise (S/N) at the edges of the galaxy, but within the region indicated in Figure~2, no spaxel has a S/N less than 5. Noise at a pixel-by-pixel basis is smoothed out by the cross-correlation techniques in any case. No regularization was imposed on any of the fitting processes.

As a first step to extract and remove emission-line contributions from the spectra, \ppxf\cite{pPXF,Cappellari17} was used to simultaneously fit the shape and kinematics of both the stellar spectra and a full set of emission lines, whose profiles were assumed to be Gaussian.  The resulting \Has emission-line flux measurements provide the tracer of ongoing star formation, since \Has luminosity $L_{\text{H}\alpha}$ is directly proportional to the local star-formation rate\cite{Kennicutt98}.

The spectra were initially logarithmically binned to allow the kinematics to be derived.  After the emission lines had been subtracted out, the remaining stellar spectra were rebinned to a linear scale and fitted using the \SL\cite{Starlight,Asari+07} code that is optimised for modelling stellar populations. (Although \SLs is explicitly designed for this task, a cross check of the results from the \SLs and \ppxfs codes confirmed very close agreement between the stellar population results obtained. This will be fully described in a separate paper by Peterken et al. (2019, in preparation).)  To reproduce the observed spectra in the fitting process, we also allowed for dust obscuration using a variable-strength Calzetti, Clayton \& Mathis reddening law\cite{CCMlaw}.

\subsection{Measuring the time offset}
The resulting output from \SLs provides the contribution of each SSP of specific age and metallicity to the spectrum at each location across the face of the galaxy.  Thus, we are now in a position to create maps showing how stars with differing properties contribute to the total light of the galaxy.  In this case, we are interested in mapping out the young stars, so we extract the contribution from all the SSPs with ages of less than $6\times10^{7}$~years.  The youngest SSP template age is $3\times10^{7}$~years, but the fitting process will attribute all younger stars to this population as well. We therefore assume that this template's `true' age is approximately $1.5\times10^{7}$~years, and all other templates' weights account for stars of their listed age. By weighting by \SL's fit's template weights, we find the mean age of the young population over the entire MaNGA field of view to be $1.9\times10^{7}$~years, with a conservative estimate of the uncertainty as $1.0\times10^{7}$~years. The peak in \Has emission occurs very soon after the onset of star-formation, and hence the temporal offset between the young population and the \Has emission in Equation~1 is $\delta\tau=19\pm10$~Myr.

\subsection{Measuring the angular offset}
Although the angular offset \ccangs between the maps of young stars and \Has emission is visually apparent in the data (see Figure~1), it is quite a subtle effect, so some care must be taken to optimise the signal when extracting it.  As a first step, we deproject the maps to face-on using kinematic centre, inclination, and position angle measurements, determined from the best-fit parameters to the gas disk kinematics, and convert the Cartesian images to polar ones, binned in radius with a step-size of $\Delta r\approx~0.16$~kpc. When the NASA-Sloan Atlas\cite{NSA} measurements are used instead for the centre, inclination and position angles (assuming the galaxy disk is intrinsically round), the results are unchanged.

\begin{figure}
\centering
\includegraphics[width=\columnwidth]{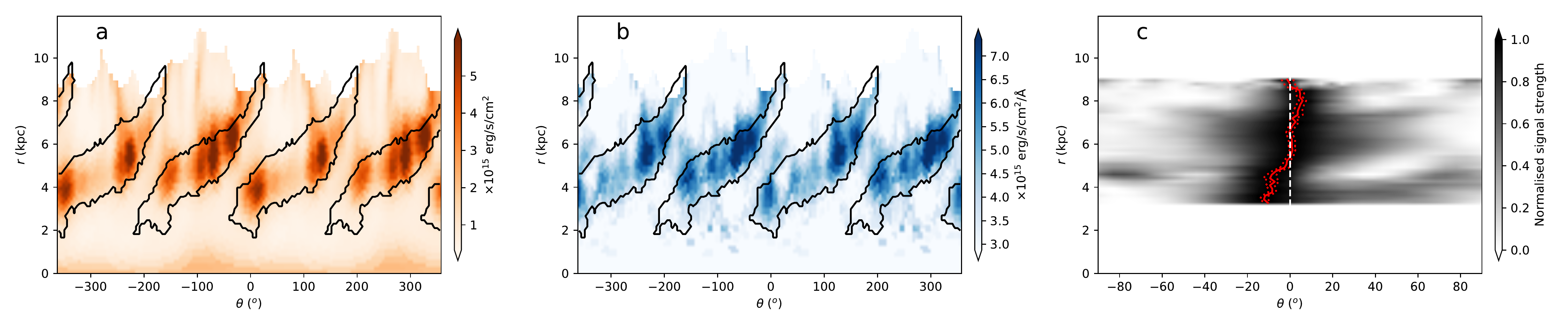}
\caption{\textbf{Cross-correlating allows us to determine the angular offset between the two tracers.} Polar-coordinate maps of the H$\alpha$ emission line and young stars are shown in (\textit{a}) and (\textit{b}). The galactic centre is at the bottom of the map, and the black outlines indicate the location of the same GZ:3D spiral arm mask used in Figure~1. The cross-correlation signal between these maps is also shown in (\textit{c}), with the cross-correlation angle \ccangs shown as a red line.}
\label{fig:Cross_correlation}
\end{figure}

For each such radius, we determine the offset between the spiral features in the \Has and young-stellar map by cross-correlating the signal in the polar maps, displayed in Figure~\ref{fig:Cross_correlation}.  The location of the cross-correlation maximum was then refined to a sub-pixel value by fitting a \nth{2}-order polynomial around the peak. The region of $r<3.2$~kpc (approximately $0.32 R_{\textrm{E}}$) in Figure~2 and is ignored when calculating \Omps since the azimuthal signal of H$\alpha$ variations here is found to be too small to reliably measure \ccang.

A conservative estimate for the uncertainty $\delta_{\Delta\theta}(r)$ in this offset can be obtained from the ratio of the full-width-at-half-maximum (FWHM) of the peak in the cross-correlation signal to the signal-to-noise ratio (SNR) of the signal. The SNR is in turn estimated as the ratio of the peak height $H$ to the standard deviation of the cross-correlation signal $\sigma_{\delta\theta(r)}$; i.e.\@
\begin{equation}
\delta_{\Delta\theta}(r) = \frac{\text{FWHM}(r)\times\sigma_{\delta\theta(r)}}{H(r)}.
\label{eq:cc_err}
\end{equation}
The radially-varying FWHM allows the value of $\delta_{\Delta\theta}(r)$ to account for the radial variation in the MaNGA beam size effects in the polar-coordinate plots. At low $r$, the beam covers a large range in $\theta$. The cross-correlation signal's peak will therefore be proportionally wider, increasing $\delta_{\Delta\theta}(r)$. At large radius, the beam will cover a small range in $\theta$, allowing us to obtain a tighter constraint on the value of \ccang.

\subsection{Measuring the angular velocity of circular orbits}
The other ingredient needed to determine the pattern speed is the angular speed of material following a circular orbit in the galaxy.  We use gas velocity measurements to determine the angular velocity of material \Oms since the very young stars this material traces will not yet have been dynamically heated from their purely circular trajectories\cite{Gerssen+97}. The MaNGA data analysis pipeline (DAP; details to be described by Westfall et al. (2018, in preparation).) provides measurements of the line-of-sight gas velocity $v_{\text{los,gas}}$ for each pixel. This analysis uses the MPL-6 version of the DAP outputs.

Using the same process as described above, the $v_{\text{gas}}$ map can be remapped into polar coordinates. At each radius, the observed line-of-sight velocity will vary sinusoidally with azimuthal angle, and a simple least-squares fit yields the amplitude of this variation at each radius, $V_\text{gas}(r)$. The angular speed can then be simply calculated as $\Omega(r) = \frac{V_\text{gas}(r)}{r\times\sin(i)}$, where $i$ is the inclination angle of the galaxy to the line of sight ($i=0$ for a face-on galaxy) derived from the galaxy's kinematics.  The error in $\Omega$ is dominated by the contribution from the uncertainty in the sinusoidal fit, and so this value is adopted.  

\subsection{Testing with an older stellar population}
\begin{figure}
\centering
\includegraphics[width=\columnwidth]{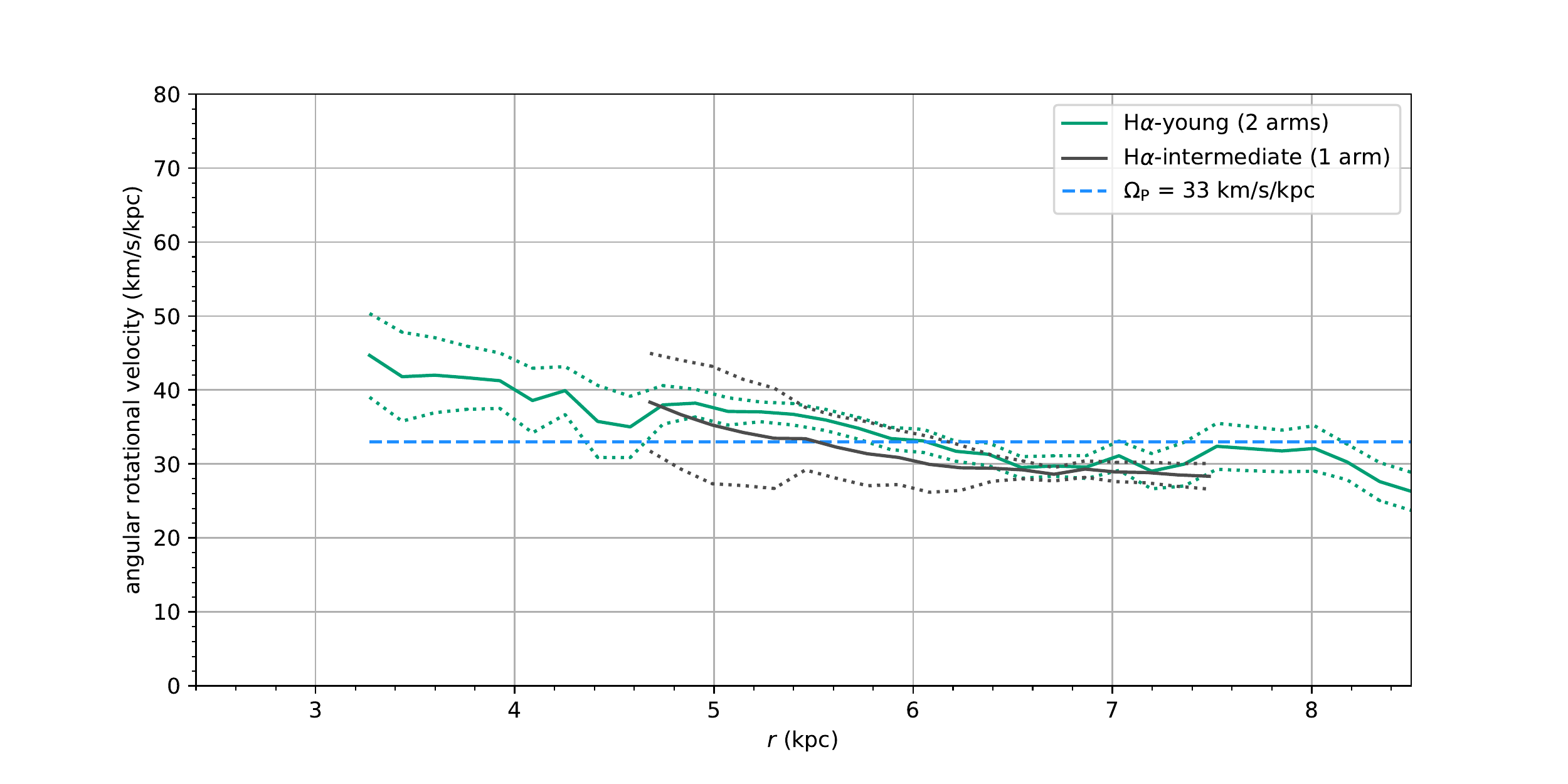}
\caption{\textbf{The result when using an older tracer is unchanged.} As Figure~2; \Omps derived in the main text (green solid) with its best-fit flat line (light blue dashed), but also showing \Omps derived from the offset between \Has and the stars of age 0.2 to 1.3~Gyr in the southern arm (grey). Dotted lines indicate estimated 1$\sigma$ uncertainties.}
\label{fig:Comparing_Indicators}
\end{figure}

If the picture established here is correct, then it should be possible to repeat the analysis using a somewhat older stellar population that will have had time to travel further from the peak of the spiral density wave.  In practice, it appears that the residual spiral feature fades very rapidly into the noise from the more general disk population (as will be described by Peterken et al. (2019, in preparation)), but we were able to extract a consistent signal from a portion of the southern spiral arm for an intermediate-age population combining the templates with ages between 0.2 and 1.3~Gyr. This population mainly comprises B- and A-type stars; adopting a luminosity-weighted age of $\approx 2\times10^{8}$~years, the results from the angular offset of these somewhat older stars, which we have shown in Figure~\ref{fig:Comparing_Indicators}, are entirely consistent with the pattern speed derived in the main text, giving some further confidence in both the method and the density wave theory.

\section{Data Availability}

Integral-field spectroscopy data of UGC~3825 is available as part of the fourteenth data release\cite{SDSSDR14} of the SDSS. The specific data that support the plots within this paper and other findings of this study are an updated version of these data will be made publicly available as part of the SDSS data release 15, which will be described in a separate paper by the MaNGA collaboration (early 2019, in preparation). In the meantime, the data used here are available from the corresponding author upon reasonable request.

\end{document}